\documentstyle[12pt,epsf,a4]{article}
\newcommand{\gsim}{\;\rlap{\lower 3.5 pt \hbox{$\mathchar \sim$}} \raise 1pt
 \hbox {$>$}\;}
\newcommand{\lsim}{\;\rlap{\lower 3.5 pt \hbox{$\mathchar \sim$}} \raise 1pt
 \hbox {$<$}\;}

\newcommand{\Lnqmone}{L_{qM_1}}
\newcommand{\Lnqmtwo}{L_{qM_2}}

\renewcommand{\thefootnote}{\fnsymbol{footnote}}
\begin{document}    

\title{\vskip-3cm{\baselineskip14pt
\centerline{\normalsize\hfill MPI/PhT/98-22}
\centerline{\normalsize\hfill hep-ph/9803313}
\centerline{\normalsize\hfill March 1998}
}
\vskip.7cm
Leptonic contribution to the effective electromagnetic
coupling constant up to three loops
\vskip.7cm
}
\author{
 M.~Steinhauser
}
\date{}
\maketitle

\begin{center}
{\it Max-Planck-Institut f\"ur Physik,
    Werner-Heisenberg-Institut,\\ D-80805 Munich, Germany\\ }
\end{center}
\vspace{.7cm}

\begin{abstract}
\noindent 
In this note the leptonic contribution to the running
of the electromagnetic coupling constant is discussed
up to the three-loop level. Special emphasize is put on
the evaluation of the double-bubble diagrams.
\end{abstract}

\vspace{.7cm}



\renewcommand{\thefootnote}{\arabic{footnote}}
\setcounter{footnote}{0}

The dominant correction to electroweak observables is
provided by the running of the electromagnetic coupling
constant from its value at vanishing momentum transfer
to high energies. The main part is delivered from the
leptonic contribution, $\Delta\alpha_{\rm lep}$,
where QED corrections to the photon
polarization function, $\Pi(q^2)$, have to be considered.
The large logarithms of the ratio between the lepton masses, $M_l$,
and the mass of the $Z$ boson, $M_Z$, is responsible for the
size of the corrections.
On the other hand the small ratio ensures that it is enough to consider an
expansion of $\Pi(q^2)$ for $q^2\gg M_l^2$. Below in addition
to the leading term we will include the first order mass corrections for the
one- and two-loop case and demonstrate that the numerical 
impact is small.

In~\cite{KueSte98} it is shown that the two-loop term for
$\Delta\alpha_{\rm lep}(M_Z^2)$ amounts to approximately $0.78\times 10^{-4}$
which though is less than $0.3$~\% of the one-loop value 
is of the same order of magnitude as the error of the hadronic
contribution, $\Delta\alpha^{(5)}_{\rm had}$~\cite{KueSte98}.
Hence, it is interesting to have a look at the three-loop contributions 
also because there for the first time quadratic logarithms of
$M_l^2/M_Z^2$ appear.

In this note the three-loop diagrams contributing to $\Pi(q^2)$,
respectively, $\Delta\alpha_{\rm lep}$, are discussed. Thereby only
the electron, muon and tau lepton are considered. Virtual quark loops
which are present at this order are not taken into account. However,
the formulae given below could in principle also be applied to this
case.

\begin{figure}[t]
 \begin{center}
 \begin{tabular}{cc}
   \leavevmode
   \epsfxsize=6.5cm
   \epsffile[142 267 470 525]{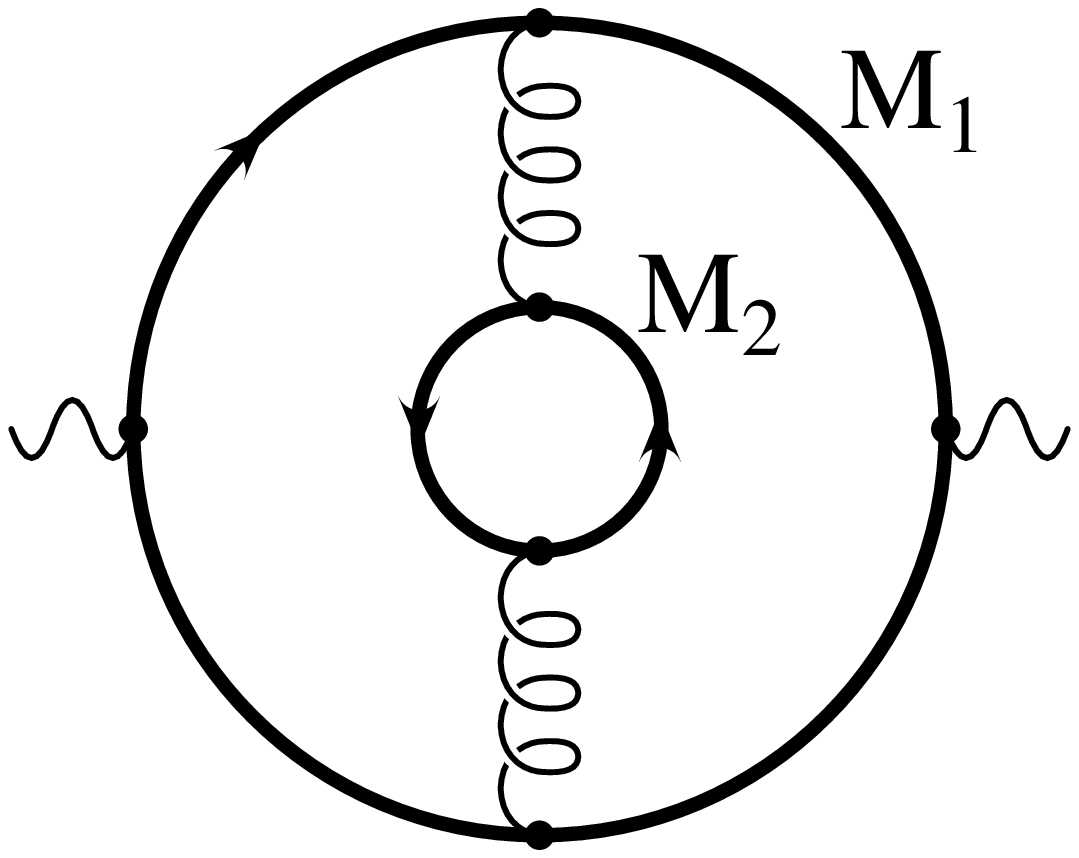}
   &
   \epsfxsize=5.0cm
   \epsffile[189 361 423 571]{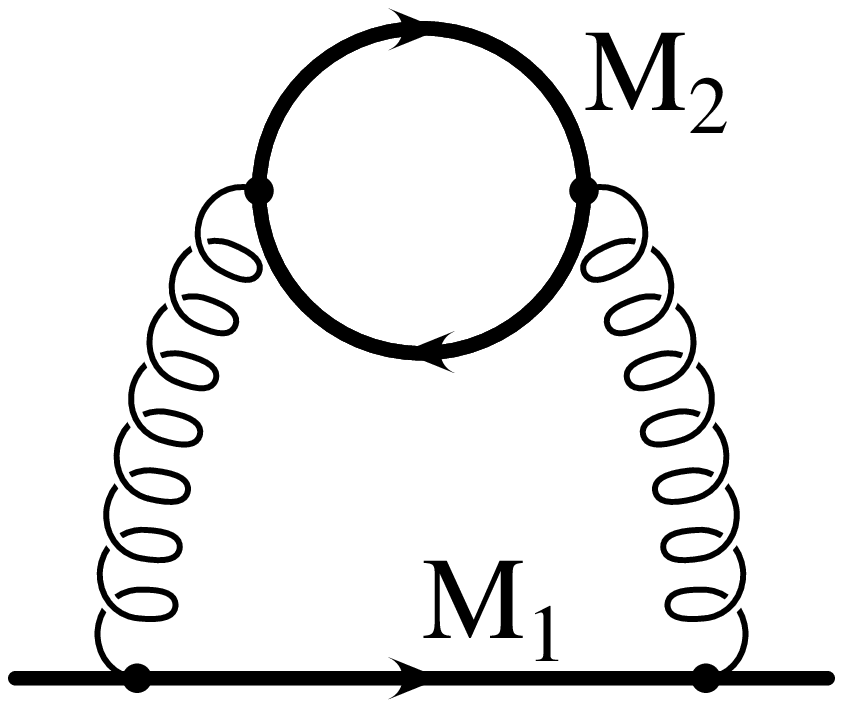}
   \\ $(a)$ & $(b)$
 \end{tabular}
\parbox{14.cm}{\small
\caption{\label{figdia}
  Double-bubble diagram $(a)$ and quark self-energy diagram
  which is needed for the computation of the mass counterterm $(b)$.
}}
 \end{center}
\end{figure}

At one- and two-loop level and for the diagrams relevant 
for the so-called quenched QED only one mass scale --- in the
following called $M_1$ --- is involved.
Concerning the double-bubble diagrams, however, in general two
mass scales, $M_1$ and $M_2$, appear, where $M_2$ is the mass
of the second fermion loop (see Fig.~\ref{figdia}$(a)$).
Having in mind the electron, muon and tau lepton 
three different cases are of practical interest: 
$(i)$ $M_1\gg M_2$, $(ii)$ $M_1=M_2$ and $(iii)$ $M_1\ll M_2$.

The quantity which enters the relation between $\alpha(0)$ and
$\alpha(M_Z^2)$ is the polarization function, $\Pi(q^2)$, evaluated for
$q^2=M_Z^2$ and normalized in such a way that $\Pi(0)~=~0$.
This means that one has to compute the current correlator
both in the limit $q^2\gg M_1^2,M_2^2$, where the masses
may be neglected, and for $q=0$.
It is convenient to perform the calculation is a first step in a
renormalization scheme where the charge and masses are defined in the
$\overline{\rm MS}$ scheme. The transformation to the on-shell
quantities is done afterwards.

The polarization function can be cast into the form
\begin{eqnarray}
\Pi(q^2) &=&
\frac{\alpha}{4\pi}\Bigg[
  \Pi^{(0)}
  +\frac{\alpha}{\pi} \Pi^{(1)}
  +\left(\frac{\alpha}{\pi}\right)^2\left(
     \Pi_{A}^{(2)}
     +\Pi_{l}^{(2)}
     +\Pi_{F}^{(2)}
     +\Pi_{h}^{(2)}
  \right)
\Bigg]
\,,
\label{eqpi}
\end{eqnarray}
where the double-bubble contributions are given by the last three
terms corresponding to the cases $(i)$, $(ii)$ and $(iii)$, respectively.
$\Pi_{l}^{(2)}$ and $\Pi_{h}^{(2)}$ depend on both masses whereas in
all other expressions only $M_1$ appears.
$\Pi^{(0)}$ and $\Pi^{(1)}$ where already computed in~\cite{KalSab55}.
The results for large external momentum can be found 
in~\cite{GorKatLar91SurSam91,BaiBro95,CheKueSte96}.
Note that in this limit the results for the three cases $(i)$, $(ii)$
and $(iii)$ are identical as mass corrections are neglected. Thus the
differences arise from the evaluation of the polarization function
for $q=0$.
The corresponding results for $\Pi_A$, $\Pi_l$ and $\Pi_F$ can be extracted 
from~\cite{BroKatTar93,BaiBro95,CheKueSte96}.
To the knowledge of the author the
case $(iii)$ was never discussed. Therefore we want to present the
result in more detail.
For the mass constellation of case $(iii)$
it is not possible to set $M_1=0$ from the very
beginning. Rather one has to apply the
so-called hard mass procedure~\cite{Smi95}
in the limit $M_1^2\ll M_2^2$.
Furthermore one encounters a two-loop mass counterterm
which has to be taken into account (see Fig.~\ref{figdia}$(b)$),
which, however, is not yet available in the literature.

Before presenting the results 
let us in a first step introduce some notation. 
Throughout the paper on-shell masses are denoted with
capital letters; $\overline{\rm MS}$ ones with small letters.
Bare masses are accompanied  with an index ``0''.
The double-bubble diagrams from case $(i)$ are multiplied with $n_l$,
the ones from $(ii)$ with $n_F$
and the ones from $(iii)$ with $n_h$.

The relation between the bare and the $\overline{\rm MS}$ mass 
reads~\cite{Tar81}:
\begin{eqnarray}
m_1^0 &\!\!=\!\!& m_1 \Bigg\{ 1 
+\frac{\bar{\alpha}(\mu^2)}{\pi} \left(-\frac{3}{4\varepsilon}\right)
\nonumber\\&&\mbox{}\qquad
+\left(\frac{\bar{\alpha}(\mu^2)}{\pi}\right)^2
 \Bigg[\frac{9}{32\varepsilon^2} -  \frac{3}{64\varepsilon}
+\left(n_l+n_F+n_h\right)
 \left(-\frac{1}{8\varepsilon^2} + \frac{5}{48\varepsilon} \right)
 \Bigg]
\Bigg\}
\,.
\end{eqnarray}
Note that the pole parts for the double-bubble diagrams are identical,
i.e., they are independent of the mass configuration ---
a special feature of MS-like schemes.
The transformation to the pole mass is performed with the help of:
\begin{eqnarray}
m_1 &=& M_1 \Bigg\{ 1 
+\frac{\bar{\alpha}(\mu^2)}{\pi} 
 \left(-1-\frac{3}{4}\ln\frac{\mu^2}{M_1^2}\right)
+\left(\frac{\bar{\alpha}(\mu^2)}{\pi}\right)^2
 \Bigg[
   \frac{7}{128}
  +\left(-\frac{15}{8}+3\ln2\right)\zeta(2)
\nonumber\\&&\mbox{}
  -\frac{3}{4}\zeta(3)
  +\frac{21}{32}\ln\frac{\mu^2}{M_1^2}
  +\frac{9}{32}\ln^2\frac{\mu^2}{M_1^2}
     + n_l\left(\frac{71}{96}
              + \frac{1}{2}\zeta(2)
              + \frac{13}{24}\ln\frac{\mu^2}{M_1^2}
              + \frac{1}{8}\ln^2\frac{\mu^2}{M_1^2}
          \right)
\nonumber\\&&\mbox{}
     + n_F\left( \frac{143}{96} 
                 - \zeta(2)
                 + \frac{13}{24}\ln\frac{\mu^2}{M_1^2}
                 + \frac{1}{8}\ln^2\frac{\mu^2}{M_1^2}
          \right)
\nonumber\\&&\mbox{}
     + n_h\left(-\frac{89}{288}
              + \frac{13}{24}\ln\frac{\mu^2}{M_2^2}
              - \frac{1}{8}\ln^2\frac{\mu^2}{M_2^2}
              + \frac{1}{4}\ln\frac{\mu^2}{M_1^2}\ln\frac{\mu^2}{M_2^2}
          \right)
 \Bigg]
\Bigg\}
\,.
\end{eqnarray}
The terms in the first three lines can be found 
in~\cite{GraBroGraSch90}. The computation is reduced to the evaluation
of two-loop on-shell integrals and can be be performed, e.g., with the
help of the program package SHELL2~\cite{FleTar92}.
The terms proportional to $n_h$
arise from the diagram pictured in Fig.~\ref{figdia}$(b)$
which has to be evaluated in the limit $M_2^2\gg M_1^2=p^2$
where $p$ is the external momentum. The bare diagram was successfully
compared with~\cite{AvdKal97}.
Using the relation
\begin{eqnarray}
\bar{\alpha}(\mu^2) &=& \alpha
\Bigg[ 1 + \frac{\alpha}{\pi} \frac{1}{3}
           \left(
                 n_l\ln\frac{\mu^2}{M_2^2}
                +n_F\ln\frac{\mu^2}{M_1^2}
                +n_h\ln\frac{\mu^2}{M_2^2}
           \right)
\Bigg]
\,,
\end{eqnarray}
also the coupling can be transformed to the on-shell scheme.
Note that the overall factor $\alpha$ in Eq.~(\ref{eqpi})
is not affected by this change of parameters.
The results for the separate contributions to $\Pi(q^2)$ then read:
\begin{eqnarray}
\Pi^{(0)} &=&
\frac{20}{9} - \frac{4}{3}\Lnqmone
+8\frac{M_1^2}{q^2}
+\ldots
\,,
\\
\Pi^{(1)} &=&
\frac{5}{6} - 4\zeta(3) - \Lnqmone
-12\frac{M_1^2}{q^2} \Lnqmone
+\ldots
\,,
\\
\Pi^{(2)}_{A} &=&
           - \frac{121}{48}
           +\left(- 5
                  + 8\ln2\right)\zeta(2)
           - \frac{99}{16}\zeta(3)
           + 10\zeta(5)
           + \frac{1}{8}\Lnqmone
+\ldots
\,,
\\
\Pi^{(2)}_{l} &=&
                 - \frac{116}{27} 
                 + \frac{4}{3}\zeta(2)
                 + \frac{38}{9}\zeta(3)
                 + \frac{14}{9}\Lnqmone
                 + \left(\frac{5}{18}
                         - \frac{4}{3}\zeta(3)\right)\Lnqmtwo
\nonumber\\&&\mbox{}
                 + \frac{1}{6}\Lnqmone^2
                 - \frac{1}{3}\Lnqmone\Lnqmtwo
+\ldots
\,,
\label{eqpi2l}
\\
\Pi^{(2)}_{F} &=&
       - \frac{307}{216} 
       - \frac{8}{3}\zeta(2)
       + \frac{545}{144}\zeta(3)
       + \left(\frac{11}{6}
             - \frac{4}{3}\zeta(3)\right)\Lnqmone
       - \frac{1}{6}\Lnqmone^2
+\ldots
\,,
\\
\Pi^{(2)}_{h} &=&
                 - \frac{37}{6} 
                 + \frac{38}{9}\zeta(3)
                 + \left(\frac{11}{6}
                        - \frac{4}{3}\zeta(3)\right)\Lnqmtwo
                 - \frac{1}{6}\Lnqmtwo^2
+\ldots
\,,
\label{eqpi2h}
\end{eqnarray}
with $\Lnqmone=\ln(-q^2/M_1^2)$ and $\Lnqmtwo=\ln(-q^2/M_2^2)$.
The dots represent subleading terms in $M_l^2/M_Z^2$.
The logarithmic dependence on $q$ of the double-bubble diagrams
coincides, the constant terms are, however, different.
It is also worth to mention that $\Pi^{(2)}_{h}$ gets independent
of $M_1$, however, only after 
the mass is transformed to the on-shell scheme.
Note also that $\Pi_l^{(2)}$ becomes dependent on $M_2$
after the result is expressed in terms of the on-shell coupling.

For the numerical evaluation of $\Delta\alpha_{\rm lep}(M_Z^2)$ 
the quantity $-\mbox{Re}(\Pi(q^2=M_Z^2))$ has to be considered taking into
account the contributions
from the electron, muon and tau lepton.
Special care has to be taken for those contributions where
two masses are involved:
In the case of the electron $\Pi^{(2)}_{l}$ is not present and
$\Pi^{(2)}_{h}$ has to be evaluated with $M_2=M_\mu$ and $M_2=M_\tau$.
For the muon $\Pi^{(2)}_{l}$ is used with $M_2=M_e$ and 
$\Pi^{(2)}_{h}$ with $M_2=M_\tau$. For the contribution from the tau
lepton $\Pi^{(2)}_{l}$ has to be evaluated with $M_2=M_e$ and
$M_2=M_\mu$.

\begin{table}[t]
\renewcommand{\arraystretch}{1.3}
\begin{center}
{\scriptsize
\begin{tabular}{|l|r|r|r|r|r|r|r|r|}
\hline\hline
$\Delta\alpha_{\rm lep}\times 10^4$
& 1-loop & 2-loop & \multicolumn{5}{|c|}{3-loop} & sum \\
\hline
& & & quenched & \multicolumn{3}{|c|}{inner lepton} & & \\
\hline
& & & $A$ & $e$ & $\mu$ & $\tau$ & sum & \\
\hline
$e$ &  174.34653 &    0.37983 &   -0.00014 &    0.00287 &
    0.00084 &    0.00025 &    0.00382 &  174.73018 \\ 
$\mu$ &   91.78419 &    0.23600 &   -0.00009 &    0.00266 &
    0.00084 &    0.00025 &    0.00366 &   92.02385 \\ 
$\tau$ &   48.05934 &    0.16034 &   -0.00007 &    0.00214 &
    0.00082 &    0.00025 &    0.00314 &   48.22282 \\ 
\hline
$ e+\mu+\tau $ &  314.19007 &    0.77617 &   -0.00030 &  --- & --- & --- &
    0.01063 &  314.97686 \\ 
\hline\hline
\end{tabular}
}
\parbox{14.cm}{\small
\caption{\label{tab1}Contributions to 
$\Delta\alpha_{\rm lep}(M_Z^2)\times 10^4$.
}}
\end{center}
\end{table}

In Tab.~\ref{tab1} the numerical results for 
$\Delta\alpha_{\rm lep}(M_Z^2)$ separated into the contributions from
the different lepton species and number of loops are listed. It can be
seen that at three-loop order for each lepton the sum of the double-bubble 
diagrams is larger by a factor of 30 to 40 as compared to the
quenched part. The total contribution of ${\cal O}(\alpha^3)$
amounts to $\approx1.4$~\% 
of the ${\cal O}(\alpha^2)$ result and is thus roughly
an order of magnitude larger than expected using the naive estimation
$\alpha/\pi\approx 2\times 10^{-3}$. This can be traced back to the
squared logarithms which are actually only present in
$\Pi_l^{(2)}$, $\Pi_F^{(2)}$ and $\Pi_h^{(2)}$.
It is also remarkable that the contributions to the double-bubble
diagrams are essentially dominated by the mass of the inner lepton
which can be understood by a closer look to 
Eqs.~(\ref{eqpi2l})--(\ref{eqpi2h}).

A comment concerning the mass corrections $M_l^2/M_Z^2$ is in order.
Obviously the most important contribution comes
from the $\tau$ lepton. It amounts at one-loop level 
to approximately $-2\times 10^{-6}$ which is of the same order of
magnitude as the one from the three-loop diagrams. The quadratic mass
corrections of order $\alpha^2$ are already two orders of magnitude smaller.
Therefore we neglect the mass corrections at the three-loop level.
In principle also terms of the form $M_1^2/M_2^2$ are present
in the correlator $\Pi_h^{(2)}$. However, they are also small and
will be neglected.

To summarize, the leptonic contribution to the effective electromagnetic
coupling constant is computed up to ${\cal O}(\alpha^3)$. At three-loop
level it turns out that the double-bubble diagrams are most important
as quadratic logarithms of the ratio $M_l^2/M_Z^2$ appear.
The total contribution to $\Delta\alpha_{\rm lep}(M_Z^2)$ amounts to $10^{-6}$.

\vspace{4mm} 

\centerline{\bf Acknowledgments} 
\medskip\noindent
I would like to thank J.H.~K\"uhn for inspiring
and valuable discussions.

\end{document}